\documentclass[10pt,letterpaper]{newFNLstyle}
\usepackage{amssymb}
\usepackage{amsmath}
\usepackage{cite}
\usepackage{graphicx}
\usepackage{ulem}
\allowdisplaybreaks

\begin{document}

\normalem

\volnumpagesyear{0}{0}{000--000}{2005}
\dates{received date}{revised date}{accepted date}

\title{Stokes' Drift and Hypersensitive Response with Dichotomous Markov Noise}

\authorsone{I. BENA$^*$}
\affiliationone{Department of Theoretical Physics, University of Geneva}
\mailingone{CH-1211, Geneva 4, Switzerland\\
$^*$Ioana.Bena@physics.unige.ch}

\authorstwo{R. KAWAI}
\affiliationtwo{Department of Physics, University of Alabama at
Birmingham}
\mailingtwo{Birmingham,  AL 35294}

\authorsthree{C. VAN DEN BROECK}
\affiliationthree{Limburgs Universitair Centrum}
\mailingthree{B-3590 Diepenbeek, Belgium}

\authorsfour{KATJA LINDENBERG}
\affiliationfour{Department of Chemistry and Biochemistry 0340 and Institute
for Nonlinear Science, University of California, San Diego}
\mailingfour{9500 Gilman Drive, La Jolla, CA 92093-0340}

\maketitle
\markboth{Stokes' Drift and Hypersensitive Response}{Bena, Kawai, Van
den Broeck, and Lindenberg}


\begin{abstract}
Stochastic Stokes' drift and hypersensitive transport driven by dichotomous 
noise are theoretically investigated.
Explicit mathematical expressions for the asymptotic probability density and 
drift velocity are derived including the situation
in which particles cross unstable fixed points. 
The results are confirmed by numerical simulations.
\end{abstract}

\section{Introduction}
\label{introduction}

The response of a Brownian particle to an external force is one of the
paradigms of statistical mechanics. 
Brownian motion is usually described as a Wiener process and a full
mathematical analysis is possible based on the solution of the corresponding
Fokker-Planck equation. However, a genuine
Gaussian white noise does not exist in real systems, and the importance
of other types of noises has long been clear.
Such noises have been studied in great detail in zero-dimensional
systems, and their specific properties are known to have a profound
influence on the behavior of these
systems~\cite{Moss:89,MFPT1}.
The effect of the color of the noise on noise-induced transitions
continues to be documented~\cite{mangioni97,mangioni00,kim98},
and has been found to be quite dramatic since it can alter the type of
transition and lead to new re-entrance phenomena. Other noise-induced
effects have also been found to be sensitive to the correlation time of
the noise, e.g., stochastic resonance~\cite{rozenfeld00},
synchronization~\cite{rozenfeld01}, and transport in Brownian
ratchets~\cite{kim98}.

The two most commonly discussed examples of colored noise are the
Ornstein-Uhlenbeck process and the dichotomous Markov
process~\cite{kampen}. The
latter has the
great advantage that analytic results can often be derived.
Nevertheless an essential technical difficulty
restricts almost all the results obtained in the past
to dynamics with no fixed points, 
or exclusively with stable fixed points,
see for example~\cite{MFPT1}. 
We have recently made progress toward overcoming
the technical difficulty that appears in the presence of unstable fixed points
by identifying the source of spurious
divergences that arise in the usual analytic approaches to the
problem~\cite{paper1,paper2}, and are now in the position to consider
these cases as well.

The purpose of this paper is to apply the general analytic procedure we
developed in Ref.~\cite{paper2} 
to two different stochastic
phenomena driven
by dichotomous Markov noise, namely Stokes' 
drift~\cite{stokesdrift} and hypersensitive
transport~\cite{ginzburg,exper}.
Section~\ref{stokes} deals with Stokes' drift as a rocking ratchet problem 
and Sec.~\ref{hyper} with hypersensitive
transport.  A short summary is presented in Sec.~\ref{conclusions}.  
Some of the lengthy calculations have been collected in the Appendices.
All the analytical results are supported by numerical simulations
of the corresponding dynamics of an ensemble of 20000 particles, with 
random initial positions, sampled at 100 different times in order to
build up the histogram of the stationary probability density and, from it,
to compute the value of the asymptotic mean velocity.

\section{Stochastic Stokes' Drift as A Rocking Ratchet}
\label{stokes}

A longitudinal wave traveling
through a viscous fluid imparts a net drift to the suspended particles, an
effect known as Stokes' drift. The deterministic effect
(that does not account for the stochastic fluctuations or perturbations in the
system) has been extensively studied in various practical contexts ranging
from the motion of tracers in  meteorology and oceanography~\cite{stokes1}
to doping impurities in crystal growth~\cite{stokes2}.
The deterministic drift has a simple intuitive explanation, namely that
the suspended particle spends a longer time in the regions of the 
wave-train where the force due to the wave acts in the direction 
of the propagation of the wave than the time it
spends in those regions where the force acts in the opposite direction.
Therefore, the particle is driven on average in the direction of 
wave propagation.
As a simple clarifying example~\cite{StokesChris}, 
consider the dynamics of
an overdamped particle forced by a traveling square wave with
velocity $v$ and wavelength $L$,
\begin{equation}
\dot{x} = f(x-vt),
\end{equation}
where
$x(t)$ is the position of the overdamped particle at time $t$, and
$f$ is the periodic forcing due to the longitudinal wave traveling
at the speed $v$ and of wavelength $L$.
Suppose that the particle is entrained
with a force $f=bv$ when in a crest part of the wave, and $f=-bv$ when
in a trough part.  The time spent in a crest 
part, $L/[2(1-b)v]$, is larger than in the though part, 
$L/[2(1+b)v]$, resulting in a net 
deterministic drift velocity $v_{0}\,=\,b^2\,v$ of the
particle.

Recent studies~\cite{stokesdrift,StokesChris,StokesDiffusion} show the
importance  
of stochastic effects on Stokes' drift. 
The thermal diffusion
of the dragged particles, as well as the application of an
external colored noise,
can markedly modify the direction and magnitude of 
the Stokes' drift velocity.
Furthermore, it has been shown that such a stochastic Stokes' drift is
equivalent to another paradigm of 
Brownian motors, the rocking ratchet~\cite{peter}.
In this
scenario, the systematic motion
acquired under the influence of an alternating zero-average stochastic  
force revolves essentially around the asymmetry of the nonlinear response
in the presence of a steady asymmetric potential~\cite{rocking}.
This mathematical equivalence between stochastic Stokes' drift and
Brownian motors hints at various potential applications of Stokes' drift, 
for instance,
transport by capillary waves, storage of light in quantum wells, 
single-electron transport in one-dimensional channels, 
optical tweezing of colloidal particles, and
diffusion of dislocations in solids, as noted 
in~\cite{marchesoni} and references therein.

In this paper we show, with an analytically solvable model,
that the characteristics of the
one-dimensional stochastic Stokes' drift are quite complex, and
that various interesting phenomena are induced, 
including enhancement of the deterministic
drift and current reversal,
when the particles are subjected to an additive
colored noise, specifically, a dichotomous noise.
The starting point is the following stochastic equation
with an additive symmetric dichotomous
perturbation:
\begin{equation}
\dot{x} = f(x-vt)+A\xi(t).
\label{eq:stokes_sde_x}
\end{equation} 
The dichotomous perturbation has an amplitude $A$ and the stochastic 
variable $\xi(t)$ takes on the values $\pm 1$ with a transition rate $k$. 
It is appropriate to assume, on physical grounds,   
that the particle cannot move faster than the wave and thus $f(y) < v$ for all 
$y$. The quantity of interest is the asymptotic drift 
velocity, $\langle\dot{x}\rangle = 
\lim_{t\rightarrow\infty}\langle x(t)\rangle/t$,
where the brackets indicate an
average over the realizations of the dichotomous noise.  Introducing a new
variable $y(t)=x(t)-vt$, one can rewrite Eq.~(\ref{eq:stokes_sde_x}) as 
\begin{equation}
\dot{y} = F(y) + A\xi(t) ,
\label{stokes4}
\end{equation} 
with a time-independent periodic force, $F(y+L) = F(y) = f(y) - v<0$. 
Equation~(\ref{stokes4}) is a variant of the general
stochastic differential equations solved in~\cite{paper1,paper2},
with a dichotomous force $F(y)\pm A$.

The behavior of the system, and the corresponding solution of 
the associated master equation for the probability density,
depend on whether or
not there are unstable fixed points in the ``$\pm$"
dynamics. We present below the results for the two simplest
cases, namely, one with no fixed points (in Sec.~\ref{stokesa}), and 
the other with two fixed points in the ``$+$"  dynamics and no fixed 
points in the ``$-$" dynamics (in Sec.~\ref{stokesb}).
 
\subsection{Systems with no fixed points}
\label{stokesa}

When 
$A>\max\left|F(y)\right|$ or
$0< A<\min\left|F(y)\right|$, 
there are no fixed points.  Then
one obtains the following expressions (see  Eqs.~(7) and~(8) in
Ref.~\cite{paper2})
for the asymptotic probability density,
\begin{equation}
   \begin{split}
       P(y) 
    =&
       \frac{\langle\dot{y}\rangle}
            {L}
       \left\{
            \left[
                 F^2(y)-A^2
            \right]
            \left[
                \exp\left(
                        \displaystyle\int_0^Ldz \frac{2k F(z)}
                                                     {F^2(z)-A^2}
                    \right)
                 -1
            \right]
       \right\}^{-1} \\
&\times
       \int_y^{y+L}dz \left[
                           F'(z)+2k
                      \right]
       \exp\left(
               -\displaystyle\int_z^y dw \frac{2k F(w)}
                                              {F^2(w)-A^2}
           \right) ,
    \end{split}
\label{stokes_case1_Px}
\end{equation}
and for the asymptotic mean velocity,
\begin{align}
\langle\dot{x}\rangle
&=v+\langle\dot{y}\rangle \nonumber 
\\
&
   \begin{aligned}
   =&\;
       v+L\left[
               \exp\left(
                       \displaystyle\int_0^L dz\;
                       \displaystyle\frac{2k F(z)}
                                         {F^2(z)-A^2}
                   \right)
                -1
          \right] \\
&\times
      \left\{
           \displaystyle\int_0^L dy
           \displaystyle\int_y^{y+L}dz
           \displaystyle\frac{F'(z)+2k}
														{F^2(y)-A^2}
           \exp\left(
                   -\displaystyle\int_z^y dw\;
                    \displaystyle\frac{2k F(w)}
                                      {F^2(w)-A^2}
               \right)
      \right\}^{-1}.
      \label{stokes10}
    \end{aligned}
\end{align}
Here $F'(y)$  denotes the derivative with respect to $y$.
The detailed analysis of these results, including the possibility of
current reversal, were already discussed in Ref.~\cite{stokesdrift}.

\subsection{Systems with asymptotic fixed points}
\label{stokesb}

If the amplitude of the noise lies in the
intermediate regime 
$\min\left|F(y)\right| < A <\max\left|F(y)\right|$,   
then the ``$+$" dynamics has at least one pair of fixed points 
in $[0,L)$.  We consider here the simplest case
in which there is only one pair of fixed points, namely,  a stable fixed point
$y_1$ with $F'(y_1)<0$, and an unstable fixed point $y_2>y_1$ with
$F'(y_2)>0$.  However, systems with several pairs of fixed points can be
treated in the same way.
According to the general discussion in Section III.B of Ref.~\cite{paper2}, 
the probability density in the interval
$[y_1,y_1+L)$ is given by
\begin{equation}
    \begin{split}
        P(y)
     =&
        \displaystyle\frac{\langle\dot{y}\rangle}
                          {L}
        \displaystyle\frac{1}
                          {\left|F^2(y)-A^2\right|}
        \displaystyle\int_{y_2}^{y}dz\;
        \text{sgn}[F^2(z)-A^2][F'(z)+2k] \\
     &\times
        \exp\left(
                -\displaystyle\int_z^y dw\;
                 \displaystyle\frac{2kF(w)}
                                   {F^2(w)-A^2}
            \right) .
    \end{split}
\label{eq:stokes_case2_Px}
\end{equation}
Equation~(\ref{eq:stokes_case2_Px}) is continuous at the unstable fixed point
$y_2$, with
\begin{equation}
\lim_{y\,\nearrow y_2} P(y)=\lim_{y\searrow y_2}P(y)=-
\displaystyle\frac{\langle\dot{y}\rangle}
                  {L}
\displaystyle\frac{2k+F'(y_2)}
                  {2A\left[k+F'(y_2)\right]}.
\end{equation}
Noting that $F'(y_2) > 0$ and $P(y_2)\geqslant 0$,
$\langle\dot{y}\rangle$ is necessarily zero or negative. 
This immediately suggests that the
direction of the net drift
$\langle\dot{x}\rangle=\langle\dot{y}\rangle+v$ can be reversed by
varying the amplitude $A$ or the transition rate $k$ of the dichotomous
perturbation.  The details of this current reversal phenomenon are discussed in
a  
later section. At the stable fixed point $y_1$ the probability density is 
continuous only for
$k/|F'(y_1)|>1$ and its value is
\begin{equation}
\lim_{y\searrow y_1}P(y)=\lim_{y\nearrow y_1}P(y)=-
\displaystyle\frac{\langle\dot{y}\rangle}
                  {L}
\displaystyle\frac{2k-|F'(y_1)|}
                  {2A\left[k-|F'(y_1)|\right]}.
\end{equation}
For $k/|F'(y_1)| \leqslant 1$, $P(y)$ is divergent at $y_1$
but integrable. 

The corresponding mean velocity is given by
\begin{equation}
    \begin{split}
        \langle\dot{x}\rangle
     =&v+L\left\{
               \displaystyle\int_{y_1}^{y_1+L} dy
               \displaystyle\int_{y_2}^{y} dz
               \displaystyle\frac{\text{sgn}\left[F^2(z)-A^2\right]
                                  \left[F'(z)+2k\right]}
                                 {\left|F^2(y)-A^2\right|}
          \right. \\
&\times
          \left.
              \exp\left(
                       -\displaystyle\int_z^y dw \;
                        \displaystyle\frac{2kF(w)}
                                          {F^2(w)-A^2}
                  \right)
          \right\}^{-1} .
    \end{split}
\label{stokes15}
\end{equation}
Note that if several waves are present, their contributions are not additive
due to the highly nonlinear dependence of $\langle\dot{x}\rangle$ on $F$.
This seems to be a general feature of the stochastic
Stokes' drift~\cite{stokesdrift,StokesChris,StokesDiffusion}, contrary to the
deterministic drift.

\subsection{The square wave}
\label{stokesc}

The above results are valid for an arbitrary form of the wave $F(y)$, 
provided that the obvious
necessary differentiability and integrability conditions are
fulfilled.
However, due to the multiple integrals in Eqs.~(\ref{eq:stokes_case2_Px})
and (\ref{stokes15}), further analytic investigation without a
simple form of $F(y)$ is difficult. Following our previous
work~\cite{stokesdrift,paper2}, we use a piecewise linear wave 
\begin{equation}
F(y)= 
   \begin{cases}
        -(1-b)v  
   &    \text{for } y \in [0,L/2-2l) \\
        -(1-b)v-\displaystyle\frac{bv}{l}(y-L/2+2l) 
   &    \text{for } y \in [L/2-2l,L/2) \\
        -(1+b)v 
   &    \text{for } y \in  [L/2,L-2l) \\
        -(1+b)v+\displaystyle\frac{bv}{l}(y-L+2l) 
   &    \text{for } y \in [L-2l,L) 
   \end{cases}   
,
\label{stokes16}
\end{equation}
with $0<b<1$, and $F(y+L)=F(y)$.
Although the analytic treatment is possible, the general solutions
are still rather complicated, and the results are relegated to \ref{stokesap}.
We focus here on the results in the limit of a square wave
$l\rightarrow 0$, for which
\begin{equation}
F(y)=
   \begin{cases}
       -F_- \equiv -(1-b)v 
   &   \text{for } y \in [0,L/2) \\[6pt]
       -F_+ \equiv -(1+b)v 
   &   \text{for } y \in [L/2,L)    
   \end{cases}
.
\label{stokes16bis}
\end{equation}
It should be noted that this limit is singular and must 
be handled with some care.

From Eq.~(\ref{stokes_case1_Px}) one obtains the probability density 
for $A > F_+$ or $0 < A < F_-$,
\begin{equation}
P(y) =
   \begin{cases}
      \displaystyle\frac{\langle\dot{y}\rangle}
            {L}
       \left[
            \displaystyle\frac{2A^2bv\left(
                                           1-e^{\phi_2}
                                     \right)
                               e^{2\phi_1 y/L}}
                              {F_+F_ - (F_-^2-A^2) \left(
                                                         e^{\phi_1+\phi_2}-1
                                                   \right)}
           -\frac{1}{F_-}
       \right] 
   &   \text{for } y \in [0,\,L/2) \\[20pt]
        \displaystyle\frac{\langle\dot{y}\rangle}
            {L}
       \left[
            \displaystyle\frac{2A^2bv\left(
                                           1-e^{\phi_1}
                                     \right)
                               e^{\phi_2(2y/L-1)}}
                              {F_+F_-(A^2-F_+^2) \left(
                                                       e^{\phi_1+\phi_2}-1
                                                 \right)}
           -\frac{1}{F_+}
       \right]
   &   \text{for } y \in [L/2,\,L)
   \end{cases} ,
\label{stokes_case1_P1}
\end{equation}
where
\begin{equation}
{\phi_{1,2}}=
\displaystyle\frac{LkF_\mp}{F_\mp^2-A^2}.
\label{phi12}
\end{equation}
Integrating Eq.~(\ref{eq:stokes_case2_Px}),  we find the probability for
$F_-<A<F_+$, 
\begin{equation}
P(y) = 
    \begin{cases}
       \begin{aligned}
           \frac{\langle\dot{y}\rangle}
                {L}
        &
           \left[
                \frac{(1-e^{\phi_1})(F_- - A)}
                     {kF_-}
                \delta_-(y-L/2)
           \right. \\
        &+
           \left.
                \frac{A\, e^{2 \phi_1 y/L}}
                         {F_-(A+F_-)}
               -\frac{1}
                     {F_-}
           \right]
        \end{aligned}
     &  \text{for }y \in (0,\,L/2) \\[24pt]
        \begin{aligned}
            \frac{\langle\dot{y}\rangle}
                 {L}
        &
            \left[
                 \frac{(1-e^{-\phi_2})(A-F_+)}
                      {kF_+}
                \delta_+(y-L/2)
            \right. \\
        &+
            \left.
                 \frac{A\, e^{2 \phi_2 (y/L-1)}}
                      {F_+(A+F_+)}
                -\frac{1}
                      {F_+}
            \right] 
        \end{aligned}
    &   \text{for } y \in (L/2,\,L)
    \end{cases}
,
\label{prob}
\end{equation}
where
$\delta_{\pm}(x)$ are the half Dirac-$\delta$
functions~\cite{footnote}.

%
\begin{figure}[t]
\begin{center}
\includegraphics{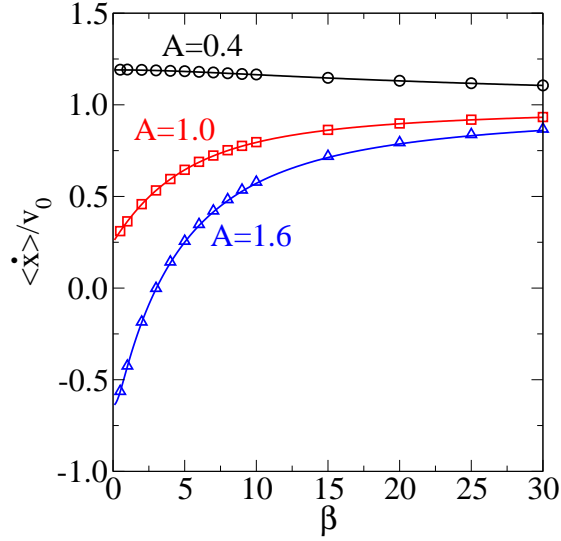}
\end{center}
\caption{The asymptotic Stokes' drift velocity  
  (normalized by its deterministic value $v_0=b^2v$)
  as a function of normalized
  transition rate $\beta=2kvL/A^2$ for three different cases. The symbols
  are the results of numerical simulations. {\it Circles:}
  $A=0.4<F_-$ (no fixed point); {\it squares:} $F_-<A=1.0<F_+$ (two fixed
  points); {\it triangles:} $A=1.6>F_+$ (no fixed  point). The solid lines
  indicate the results of theory. The values of the other
  parameters are $v=1$, $b=0.5$,  and $L=1$.  The direction of
  the drift velocity remains the same for $A=0.4$ and $A=1.0$, but
  is reversed for $A=1.6$.}
\label{valphastokes}
\end{figure}

Correspondingly, the  drift velocity for $A > F_+$ or $0 < A < F_-$
is given by
\begin{equation}
{\langle\dot{x}\rangle}
=v_0\;\displaystyle\frac{1-e^{\phi_1+\phi_2}-\displaystyle\frac{2A^2v}
                                                          {LkF_+F_-}
                     (1-e^{\phi_1})(1-e^{\phi_2})}
                   {1-e^{\phi_1+\phi_2}-\displaystyle\frac{2A^2v_0}
                                                          {LkF_+F_-}
                     (1-e^{\phi_1})(1-e^{\phi_2})} ,
\label{eq:stokes_sqw1_xdot}
\end{equation}
and for $F_-<A<F_+$ by
\begin{equation}
\langle\dot{x}\rangle
= v - \left[
           \frac{1}
                {v-v_0}
          -\displaystyle\frac{(1-e^{\phi_1})(F_- - A)^2}
                             {2kLF_-^2}
          +\displaystyle\frac{(1-e^{-\phi_2})(F_+ - A)^2}
                             {2kLF_+^2}
   \right]^{-1},
\label{eq:stokes_sqw2_xdot}
\end{equation}
where, recall, $v_0=b^2 v$ is the deterministic value of
the Stokes' drift velocity for the square wave.
We mention some limiting cases of interest.  
(i) The limit $A\rightarrow 0$ or $k\rightarrow \infty$ leads to
the deterministic Stokes' velocity,
$\langle\dot{x}\rangle={v_{0}}$. 
(ii) The drift velocity derived
in Ref.~\cite{StokesChris} is recovered in the white noise limit $A\rightarrow
\infty$, $k\rightarrow\infty$, with a finite $D=A^2/2k$.
(iii)
The quenched-noise limit $k\rightarrow 0$ describes the system in which
half of the particles, chosen at random, are subjected to a constant
external forcing $+A$ and the other half to forcing
$-A$. When there are no fixed points, the mean velocity in this limit is 
\begin{equation}
{\langle\dot{x}\rangle}={v_{0}}\;\left(\frac{v^2}
                                         {v^2-A^2}\right) .
\end{equation}
This expression clearly indicates the existence of the flux reversal
suggested earlier, since the drift velocity is negative for sufficiently
large amplitude $A$ of the forcing, and positive for small $A$'s.
When there are fixed points in the ``+" dynamics,
Eq.~(\ref{eq:stokes_sqw2_xdot}) 
leads to a mean velocity 
\begin{equation}
\langle\dot{x}\rangle = \frac{F_+^2 + F_-^2 - 2 A^2}
                             {4(v+A)} .
\end{equation}
In this limit, the current is reversed
at $A=\sqrt{(F_+^2+F_-^2)/2}=\sqrt{1+b^2}\;v$.

Figure~\ref{valphastokes} shows the drift velocity, 
Eqs.~(\ref{eq:stokes_sqw1_xdot}) and (\ref{eq:stokes_sqw2_xdot}), as
a function of the 
normalized transition rate, $\beta=2kvL/A^2$.  The plots of the analytic
results 
are in good agreement with the Monte Carlo simulation of the original
stochastic 
differential equation (\ref{eq:stokes_sde_x}).  When the transition rate is
high ($\beta \gg 1$), the effect of the dichotomous noise is averaged out and
the drift velocity approaches the deterministic limit $v_0$ regardless 
of the noise amplitude.  Interesting
phenomena, namely, negative drift velocity and current reversal, take place
only when the transition rate is sufficiently small, i.e.,  $\beta \lesssim 1$. 

%
\begin{figure}
\begin{center}
\includegraphics[]{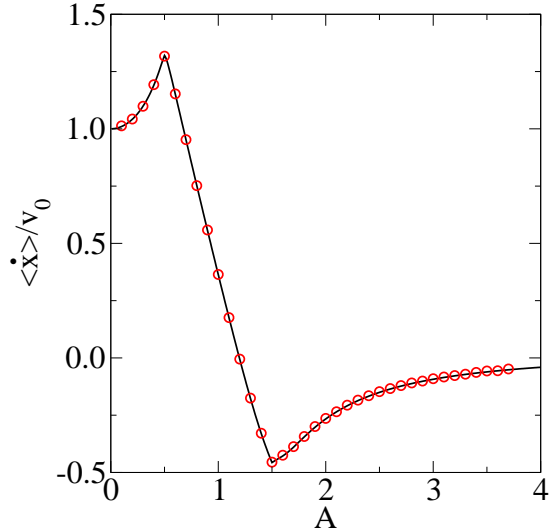}
\end{center}
\caption{The asymptotic Stokes' drift velocity 
  (normalized by its deterministic value $v_0=b^2v$)
  as a function of the noise
  amplitude $A$. The solid line represents theory and the circles show
the results
  of numerical simulations. The values of the parameters are $v=1$, $b=0.5$,
  $k=1$, and $L=1$.} 
\label{vAstokes}
\end{figure}
%
\begin{figure}
\begin{center}
\includegraphics[]{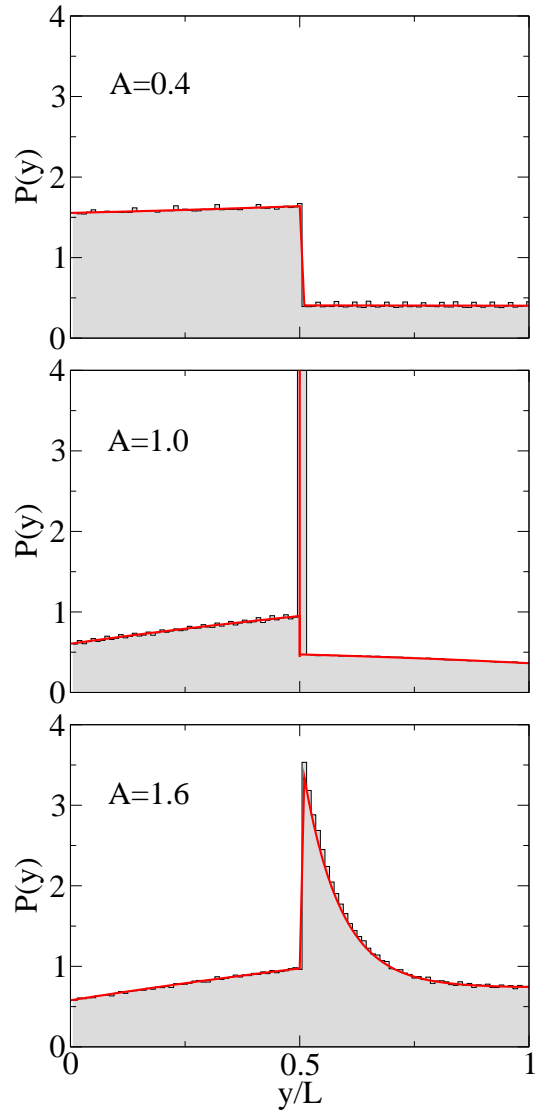}
\end{center}
\caption{The profile of the probability density $P(y)$ for three
values of the amplitude $A$ of the dichotomous noise. The
first ($A<F_-$) and the
last panel ($A>F_+$) correspond to the
regime with no fixed points, while the second panel ($F_-<A<F_+$) corresponds
to the 
presence of
two fixed points. Solid lines indicate the analytic results and 
shaded histograms show the results of numerical simulation. The values 
of the parameters are 
$v=1$, $b=0.5$, $k=1$, and $L=1$.  In the simulation, the dynamics of 20000 
particles is sampled at 100 different times.}
\label{fig:stokes_Px}
\end{figure}

Figure~\ref{vAstokes} shows the drift velocity,
Eqs.~(\ref{eq:stokes_sqw1_xdot}) and
(\ref{eq:stokes_sqw2_xdot}), in the three different regimes
corresponding to low ($A<F_-$) and high ($A>F_+$) noise amplitudes with no
fixed point, and  
intermediate ($F_-<A<F_+$) noise amplitudes with a pair of fixed points.
The agreement between 
analytic results and simulation is again very good.
At low noise amplitudes the
drift velocity is positive and increases 
with increasing noise amplitude, i.e., noise enhances 
Stokes' drift.  The noise-induced enhancement reaches its maximum at $A=F_-$
and decreases above this noise amplitude.   
In general, noise-induced phenomena
disappear when the noise becomes too large.   However, the decrease at the
intermediate noise is not
the destruction of the Stokes' drift due to large fluctuations of the
particle velocity.   The sudden change with 
discontinuity in the first derivative of the drift velocity suggests 
the appearance of a bifurcation in the dynamics.  
When $F_-<A<F_+$, the system has a stable fixed
point in the ``+" dynamics and some particles become stuck at the fixed
point until the noise switches.  The actual drift takes place only when the
system is in the ``--"  dynamics, corresponding to a smaller or even
negative value of $\dot{x}$.  The negative drift reaches its maximum at $A=F_+$
where the fixed points disappear.  For $A>F_+$, the ``$\pm$''
dynamics correspond to drift velocities in opposite directions, and therefore
the net drift is reduced. Finally, as expected, Stokes' drift is asymptotically
destroyed as the noise amplitude increases to infinity.

Figure~\ref{fig:stokes_Px} illustrates the probability densities
Eqs.~(\ref{stokes_case1_P1}), 
(\ref{prob}) for three
different amplitudes of the dichotomous noise. For comparison,
the numerical solution of
the stochastic differential equation (\ref{eq:stokes_sde_x}) is also shown.
The agreement between the present analytic theory and the computer
simulation is nearly perfect.
Due to the discontinuity of the square wave $F(y)$, Eq.~(\ref{stokes16}), 
the probability densities are
discontinuous at $y=0$ and $y=L/2$ for all cases.
Moreover, for $A=1.0$ the
system has fixed points and the probability density becomes $\delta$-singular at the
stable fixed point.  The probability densities are also asymmetric in all cases.
For 
$A=0.4$, the density is higher at $y \in (0,L/2)$ than
at $y \in (L/2, L)$.  On the
other hand, the situation is the other way around for $A=1.6$.  This difference
causes the current reversal.

\section{Hypersensitive Transport}
\label{hyper}

Generally and somewhat loosely speaking, the term ``hypersensitive
transport" refers to a
highly nonlinear directed response of a nonequilibrium, noisy system to
a small systematic external forcing.
This phenomenon was discovered rather recently, and received a great
deal of 
theoretical~\cite{paper1,ginzburg} and experimental~\cite{exper}
attention.

One of the simplest models exhibiting this novel phenomenon
describes an overdamped particle whose dynamics switches dichotomically
between a symmetric potential and its negative. When the symmetry of the
system is slightly broken
by a small directed external force, the system responds highly
nonlinearly by exhibiting a ``giant" systematic particle drift.
Let us thus 
consider the following stochastic dynamics 
with a multiplicative noise~\cite{paper1}:
\begin{equation}
\dot{x}=F+\xi(t)v(x).
\label{hyper1}
\end{equation}
The dichotomous noise $\xi(t)=\pm1$ has, as before, transition
rate $k$. $F>0$ represents a constant external force,  
and $v(x)$ is a given symmetric ``substrate" force profile that is assumed 
to be periodic, $v(x+L)=v(x)$.  Some preliminary results on this model were
reported in Ref.~\cite{paper1}. Here we present
more (and more detailed) results than in our earlier work.

%
\begin{figure}
\begin{center}
\includegraphics[width=2.8in]{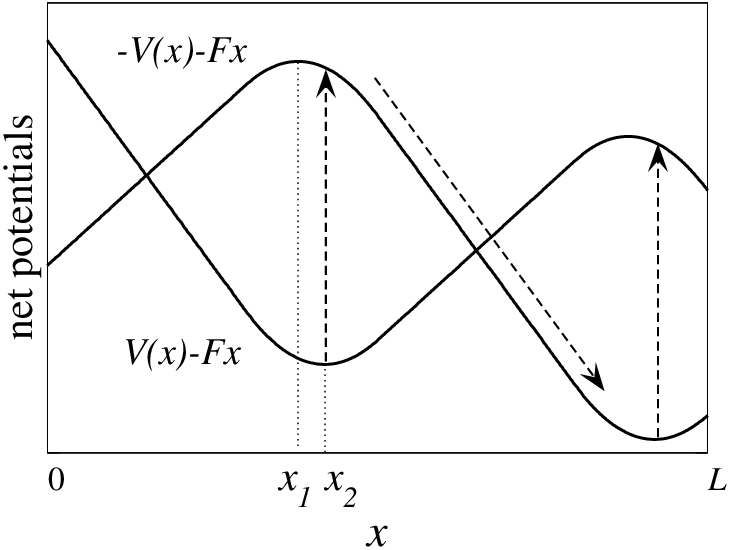}
\end{center}
\caption{
The net potentials $\mp V(x) -Fx$.  $x_1$ and $x_2$ 
represent unstable and stable fixed points, respectively.  The substrate
potential is defined as $V(x) = -\int^x v(x') dx'$.  When the switching rate
between these two potentials is sufficiently small, particles accumulate at a
stable fixed point $x_2$, bypassing the unstable fixed point $x_1$.  When the
potential switches, the particles are repelled by the unstable fixed point
$x_1$ toward a new stable fixed point.  Even when the external force $F$ is
very small, the particles are able to avoid backward drift.
}
\label{potentials}
\end{figure}

\subsection{No fixed points}
\label{hypera}

When the external force is sufficiently large 
[$F^2-v^2(x)\neq 0$ for any $x\in [0,L)$], 
there is no fixed point in either dynamics, and
one obtains the asymptotic probability density
\begin{equation}
P(x)=
\frac{\langle\dot{x}\rangle}
     {LF}
\left\{
     1 + \frac{v(x)\displaystyle\int_x^{x+L}dz\;
               v'(z)\exp\left[
                            -\int_z^x
                             dw \frac{2kF}
                                  {F^2-v^2(w)}
                         \right]}
             {[F^2-v^2(x)]
              \left\{
                   \exp\left[
                            \displaystyle\int_0^L
                            dz \frac{2kF}
                                 {F^2-v^2(z)}
                       \right]
                   -1
             \right\}}
\right\} ,
\label{hyper3}
\end{equation}
and, through the normalization of $P(x)$, the mean asymptotic velocity
\begin{equation}
\frac{\langle\dot{x}\rangle}
     {F}
=
\left\{
      1+\frac{\displaystyle\int_0^Ldx\,\frac{v(x)}
                                            {F^2-v^2(x)}
              \int_x^{x+L}dz\;
              v'(z)\exp\left[
                           -\int_z^x
                            dw \frac{2kF}
                                 {F^2-v^2(w)}
                       \right]}
             {L\left\{
                    \exp\left[
                             \displaystyle\int_0^L
                             dz \frac{2kF}
                                  {F^2-v^2(z)}
                       \right]
                    -1
              \right\}}
\right\}^{-1}.
\label{hyper4}
\end{equation}

\subsection{Asymptotic dynamics with fixed points}
\label{hyperb}

For simplicity and without the loss of any relevant point of
the method, we take $v(x)$ to be a continuously decreasing 
function in $[0,L/2]$
and symmetric about $L/2$, $v(x+L/2)=-v(x)$. This implies that
$P(x+L/2)=P(x)$, so that we can limit 
our analysis to half a period. 
When the external force is weak, the particle moves 
alternately in the two ``net potentials"
represented schematically in Fig.~\ref{potentials}. 
In this simple case
the equation  $F^2-v^2(x)=0$ has only two solutions in $[0,L/2]$, 
$x_1$, corresponding to an unstable fixed point in the ``$-$''
dynamics [$F=v(x_1)$, $v'(x_1)<0$], and $x_2$, a stable fixed point in the
``$+$'' dynamics [$F=-v(x_2)$, $v'(x_2)<0$], with $x_2>x_1$. These 
fixed points are the local extrema of the net potentials.

According to the discussions in~\cite{paper1,paper2}, the
physically acceptable solution in the interval
$[x_2-L/2,x_2]$ is given by
\begin{equation}
   \begin{split}
       P(x) =&
       \displaystyle\frac{\langle\dot{x}\rangle}
                         {LF}
       \left\{
              1+\displaystyle\frac{v(x)}
                                  {|F^2-v^2(x)|}
             \displaystyle\int_{x_1}^xdz\;
             \text{sgn}\left[
                             F^2-v^2(z)
                      \right]v'(z)
       \right.  \\
&\times
       \left.
           \exp\left[
                    -\displaystyle\int_z^x
                    dw\frac{2kF}
                         {F^2-v^2(w)}
               \right]
       \right\} ,
   \end{split}
\label{hyper6}
\end{equation}
which extends by periodicity to the whole $x$-axis.
At the unstable fixed point $x_1$, the probability 
density is continuous and its value is given by
\begin{equation} 
\lim_{x\searrow x_1}P(x)
=\lim_{x\nearrow x_1}P(x)
=\frac{\langle\dot{x}\rangle}
      {LF\left\{
              1-1/[{2(k/|v'(x_1)|+1)}]
         \right\}} .
\end{equation}
Note that the stable fixed points $x_2-L/2$ and $x_2$ are located at
the ends of the chosen interval and 
there is no other stable fixed point within this interval.  
Therefore, the probability density
is continuous inside the interval.
However, it can be singular  at the stable fixed points.
When the transition rate is sufficiently large [$k/|v'(x_2)|>1$],
there is not enough time for the particles to reach and to
accumulate at the stable fixed points,
and thus $P(x)$ is continuous  $x_2$ and $x_2-L/2$ as follows:
\begin{equation}
\lim_{x\nearrow x_2}P(x)
=\lim_{x \searrow (x_2-L/2)}P(x)
=\frac{\langle\dot{x}\rangle}
      {LF\left\{
              1+1/[{2(k/|v'(x_2)|-1)}]
         \right\}}.
\end{equation}
For low transition rates [$k/|v'(x_2)| \leqslant 1$], the majority of particles 
become stuck near the stable fixed points for a long time, thus
resulting in the divergence of $P(x)$ at $x_2$ and $x_2-L/2$.
These divergences
cause a highly nonlinear conductivity of the system, as discussed later.

From the normalization of $P(x)$, the average velocity is obtained as
\begin{equation}
    \begin{split}
       \frac{\langle\dot{x}\rangle}
            {F}
     =&
       \left\{
              1+\frac{2}{L} 
             \displaystyle\int_{x_2-L/2}^{x_2}dx\;
             \frac{v(x)}
                  {|F^2-v^2(x)|} 
       \right. \\
     &\times
       \left. 
           \displaystyle\int_{x_1}^xdz\;
           \text{sgn}\left[
                           F^2-v^2(z)
                     \right]v'(z)
           \exp\left[
                   -\displaystyle\int_z^x
                    dw\frac{2kF}
                         {F^2-v^2(w)}
               \right]
       \right\}^{-1} .
    \end{split}
\label{hyper11}
\end{equation}

Although exact, the above results are
still too complicated to gain a clear physical picture
of the behavior of the system. 
We therefore turn to a particular shape of the velocity profile
$v(x)$ that simplifies the evaluation of the above functional
expressions.

\subsection{Piecewise linear internal force}
\label{hyperc}

We consider a piecewise linear ``substrate" force
\begin{equation}
v(x)=
   \begin{cases}
       v_0, 
   &   \text{for } x \in [0,L/2-2l) \\[3pt]
       v_0 \left[
                 L/(2l)-1- x/l
           \right],
   &   \text{for } x \in[L/2-2l,L/2) \\[3pt]
      -v(x\,-L/2) 
   &   \text{for } x \in [L/2,L)
   \end{cases}
\label{hyper12}
\end{equation}
with $l\leqslant L/4$ and the periodicity condition
$v(x+L)=v(x)$. It is convenient to introduce the 
following dimensionless
variables:
\begin{equation}
f=F/v_0,\qquad \alpha=lk/v_0, \qquad \Gamma=4l/L \quad (0<\Gamma<1),
\label{hyper13}
\end{equation}
and to work with the function 
\begin{align}
T(x)
 =& \displaystyle\frac{1}
                      {v_0}
    \displaystyle\int_0^xdz\;\text{sgn}\left[
                                             F^2-v^2(z)
                                       \right]
     v'(z) \exp\left[
                   \displaystyle\int_0^z dw \frac{2kF}
                                              {F^2-v^2(w)}
               \right] \nonumber \\[14pt]
 =&
    \begin{cases}
        0
     & \text{for } x\in [0,L/2-2l)\\[6pt]
        \begin{aligned}
          &\left|\displaystyle\frac{f+1}
                                   {f-1}
           \right|^{\alpha}
           \exp \left[
                    -\displaystyle\frac{4\alpha f(1-\Gamma)} 
                                       {(1-f^2)\Gamma}
                \right] \\[8pt]
          &\times
           \displaystyle\int_1^{\chi} ds\;
           \text{sgn}(f^2-s^2)
           \left|\displaystyle\frac{f-s}
                                   {f+s}
           \right|^{\alpha} 
        \end{aligned}
     &  \text{for } x \in [L/2-2l,L/2)
    \end{cases}
,
\label{eq:hyper_Tx}
\end{align} 
where $\chi = L/(2l)-1-x/l$.

In the absence of fixed points ($f>1$), one obtains the stationary probability density 

\begin{equation}
P (x)= 
   \begin{cases}
       \begin{aligned}
           \displaystyle\frac{\langle\dot{x}\rangle}
                             {LF}
        &
           \left\{
                  1-\frac{T(L/2)}
                         {(f^2-1)(\Delta+1)}
           \right. \\[3pt]
        &  \times
           \left.
               \exp\left[
                       -\frac{8\alpha f}
                             {(f^2-1)\Gamma}
                        \frac{x}{L}
                   \right]
           \right\}
       \end{aligned}
    &  \text{for } x \in [0,L/2-2l) 
\\[30pt]
       \begin{aligned}
           \displaystyle\frac{\langle\dot{x}\rangle}
                             {LF}
          &
           \left\{
                  1+\displaystyle\left(\frac{f+1}
                                            {f-1}
                                 \right)^{\alpha}
           \right. \\[3pt]
          &\times
           \left.
               \displaystyle\frac{\chi
                                  \left[
                                        T(x)-T(L/2) /(\Delta+1)
                                  \right]}
                                 {\Delta(f-\chi)^{1+\alpha}
                                        (f+\chi)^{1-\alpha}}
           \right\}
       \end{aligned}
     & \text{for } x \in [L/2-2l,L/2)
    \end{cases} ,
\label{eq:hyper_case1_Px}
\end{equation}
where we have introduced the short-hand notation
\begin{equation}
\Delta=\displaystyle\left|\frac{f+1}{f-1}\right|^{2\alpha}
\exp\left[\frac{4\alpha f(1-\Gamma)}{(f^2-1)\Gamma}\right].
\label{hyper18}
\end{equation}
The corresponding mean velocity is given by
\begin{equation}
    \begin{split}
       \displaystyle\frac{\langle\dot{x}\rangle}
                         {F}
    =&
       \Biggl\{
               1-\displaystyle\frac{\Gamma T(L/2)}
                           {4\alpha f(\Delta+1)}
              \left[
                    1-\Delta^{-1}
              \left(
                   \displaystyle\frac{f+1}
                                     {f-1}
              \right)^{2\alpha}
              \right]
             +\displaystyle\frac{\Gamma}
                                {2\Delta}
              \left(
                   \displaystyle\frac{f+1}
                                     {f-1}
              \right)^{\alpha}
       \Biggr.  \\
    &\times
       \Biggl.
       \displaystyle\int_{-1}^1dt\;
       \displaystyle\frac{t\left[
                                 T(L/2-l(t+1))-T(L/2)/(\Delta+1)
                           \right]}
                         {(f-t)^{1+\alpha}(f+t)^{1-\alpha}}
       \Biggr\}^{-1}.
    \end{split}
\label{eq:hyper_case1_xdot}
\end{equation}

When $0<f<1$ the dynamics has two fixed points,  
an unstable one at
$x_1=L/2-l(1+f)$  and a stable one at $x_2=L/2-l(1-f)$.  
The situation is then more
complicated but still analytically tractable. 
The probability density is  written separately for three different regions:
\begin{equation}
P(x)=
    \begin{cases}
        \displaystyle\frac{\langle\dot{x}\rangle}
                          {LF}
        \left\{
               1-\displaystyle\frac{T(x_1)}
                                   {1-f^2}
              \exp\left[
                       \displaystyle\frac{8\alpha f}
                                         {(1-f^2)\Gamma}
                       \displaystyle\frac{x}
                                         {L}
                  \right]
        \right\}
    &   \text{for } x\in [0,L/2-2l) \\[14pt]
        \begin{aligned}
            \displaystyle\frac{\langle\dot{x}\rangle}
                              {LF}
        &   \left\{
                   1+\displaystyle\left(
                                        \frac{1+f}
                                             {1-f}
                                  \right)^{\alpha}
            \right. \\
        &\times 
            \left.
                \displaystyle\frac{\chi\left[
                                             T(x)-T(x_1)
                                       \right]}
                                  {\Delta \left|
                                               f-\chi
                                          \right|^{1+\alpha} 
                                          \left|
                                              f+\chi
                                          \right|^{1-\alpha}}
            \right\}
        \end{aligned}
     &  \text{for } x\in [L/2\,-\,2\,l,\,x_2) \\[28pt]
        \begin{aligned}
            \displaystyle\frac{\langle\dot{x}\rangle}
                              {LF}
        &   \left\{
                   1+\displaystyle\left(
                                       \frac{1+f}
                                            {1-f}
                                  \right)^{\alpha}
            \right. \\
        &\times
            \left. 
                \displaystyle\frac{\chi\left[
                                             T(x)+\Delta T(x_1)-T(L/2)
                                       \right]}
                                  {\Delta\left|f-\chi
                                         \right|^{1+\alpha} 
                                         \left|f+\chi
                                        \right|^{1-\alpha}}
            \right\}
        \end{aligned}
    &  \text{for } x \in [x_2,L/2)
    \end{cases}
.
\label{eq:hyper_case2_Px}
\end{equation}
This probability density is continuous at the unstable fixed point
$x_1$,
\begin{equation} 
\lim_{x\searrow x_1}P(x)
=\lim_{x \nearrow x_1}P(x)
=\frac{\displaystyle{\langle\dot{x}\rangle}}
                    {LF\left\{
                            1-1/[2(\alpha+1)]\right\}} .
\end{equation}
At the stable fixed point $x_2$, $P(x)$ is continuous for
$\alpha>1$, with
\begin{equation}
\lim_{x\searrow x_2}P(x)
=\lim_{x \nearrow x_2}P(x)
=\frac{\displaystyle{\langle\dot{x}\rangle}}
                    {LF\left\{
                            1+1/[2(\alpha-1)]\right\}},
\end{equation}
but divergent and integrable for $\alpha \leqslant 1$.
The average velocity as usual follows from the
normalization of $P(x)$,
\begin{equation}
    \begin{split}
        \displaystyle\frac{\langle \dot{x} \rangle}
                          {F} 
    =&
        \Biggl\{
                1-\frac{\Gamma T(x_1)}
                       {4\alpha f}
               \left[
                    \left(
                         \displaystyle\frac{1+f}
                                           {1-f}
                    \right)^{2\alpha}
                    \Delta^{-1}-1
               \right] + \frac{\Gamma(1+f)^{\alpha}}
                              {2\Delta(1-f)^{\alpha}}
       \Biggr. 
\\
    &\times
       \Biggl.
            \biggl\{
                    \displaystyle\int_{-1}^{1}dt\;
                    \displaystyle\frac{t\bigl[
                                       T[L/2-l(t+1)]-T(x_1) 
                                        \bigr]}
                                      {\left|t-f\right|^{1+\alpha}
                                       \left|t+f\right|^{1-\alpha}}
                    + \left[
                            T(x_1)(1+\Delta)-T(L/2)
                      \right]
            \biggr.
      \Biggr.
\\
   &\times
      \Biggl.
           \biggl.
                \displaystyle\int_{-1}^{-f}dt\; 
                \displaystyle\frac{t}
                                  {\left|t-f\right|^{1+\alpha}
                                   \left|t+f\right|^{1-\alpha}}
           \biggr\}
     \Biggr\}^{-1}.
   \end{split}
\label{eq:hyper_case2_xdot}
\end{equation}
The integrals in Eqs.~(\ref{eq:hyper_Tx}), (\ref{eq:hyper_case1_xdot}), and
(\ref{eq:hyper_case2_xdot}) 
cannot be evaluated in closed analytic form except for a few specific
values of $\alpha$.  
As an illustration, in~\ref{hyperparticular} we present closed
analytic results for $\alpha=1/2$.

%
\begin{figure}
\begin{center}
\includegraphics[]{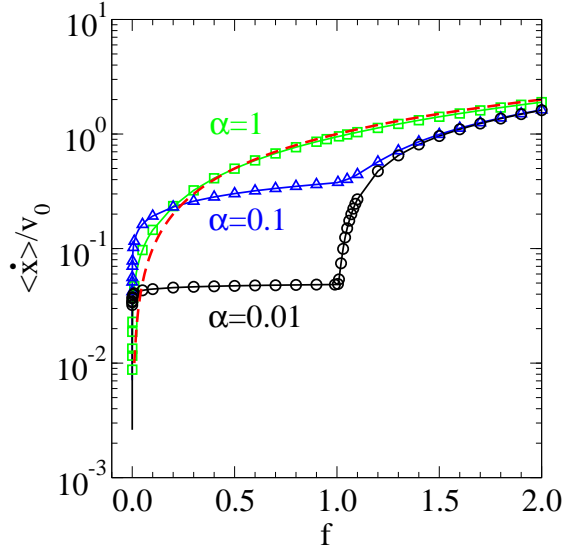}
\end{center}
\caption{The mean asymptotic velocity as a function of the
applied force $f$ for different values of the parameter $\alpha$. 
The other parameters are $v_0=1$, $\Gamma=0.4$, and $L=1$. 
The symbols represent the results of numerical simulations.
The solid lines are the result of theory
and the dashed line indicates the linear response.}
\label{vfhyper}
\end{figure}

%
\begin{figure}
\begin{center}
\includegraphics[width=7.5cm]{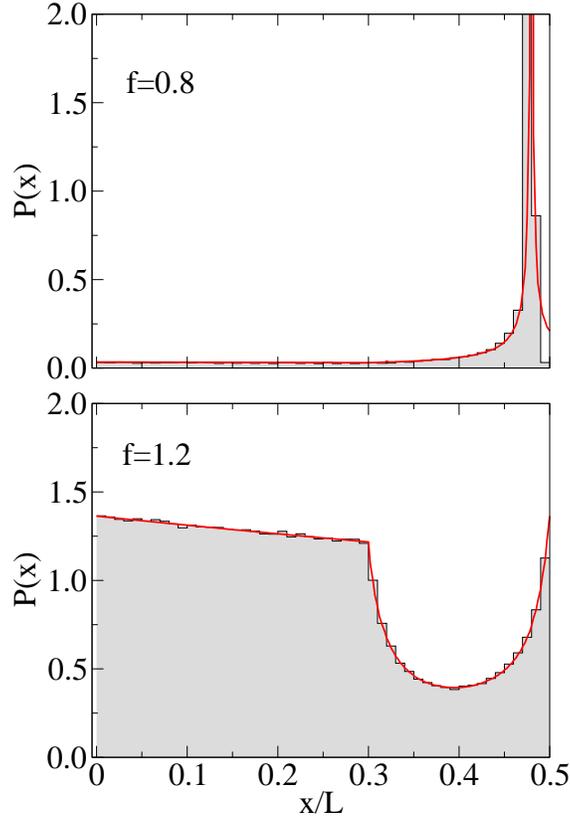}
\end{center}
\caption{The profile of the probability density $P(x)$ for two values of the
applied force $f$. The first corresponds to the
regime with fixed points, and the second to the regime with no fixed points.
Solid lines indicate the analytic
results and shaded histograms show the results of numerical simulations.
The values of the other parameters are 
$v_0=1$, $\alpha=1$, $\Gamma=0.4$, and $L=1$. In the simulations, the dynamics
of 20000 particles is sampled at 100 different times.}
\label{pxhyper}
\end{figure}

\subsection{Response Properties}

Equations~(\ref{eq:hyper_case1_xdot}) and (\ref{eq:hyper_case2_xdot}) are
numerically evaluated and 
plotted in Fig.~\ref{vfhyper}  as a function of the external force $f$ for
different values of $\alpha$. The results are in good agreement with computer
simulations. 
The drift velocity shows highly nonlinear responses to the external force
depending on the transition rate of the dichotomous noise, as illustrated in
Fig.~\ref{vfhyper}.  There are four asymptotic regimes of interest. 

(i) {\it Linear Response Regime I}: When the deterministic external force $f$
is much larger 
than the fluctuating force ($f \gg 1$), the effect of the fluctuating force
is negligible and a linear response, $\langle\dot{x}\rangle/v_0 = f$,
is expected
for any transition rate.  Indeed, all curves in Fig.~\ref{vfhyper} approach 
the linear response curve as $f$ increases.

(ii) {\it Linear Response Regime II}:  When the transition rate is
very high ($\alpha \gg 1$), the fluctuating force is averaged out and only the
deterministic external force effectively drives the particles.
Therefore, a linear response, $\langle\dot{x}\rangle/v_0 = f$,
is again expected for $\alpha \gg 1$.  Figure~\ref{vfhyper} shows that the drift 
velocity for $\alpha=1$ is already very close to this linear response limit.
 
(iii) {\it Adiabatic Regime:} Since we are
interested in nonlinear response, we focus on the cases with small
$\alpha$ and small $f$.  When the transition rate is sufficiently
small (adiabatic limit), the system remains in one of
the $\pm V(x)$ potentials for
a long time.  The
particles eventually reach 
a minimum of the net potential and wait there until the potential flips.
When the dichotomous noise switches its value, the particles suddenly find
themselves near the maximum of the net potential as illustrated in
Fig.~\ref{potentials}.  
They move down to a new
potential minimum and wait again for the next potential flip.
A typical time for them to
escape from the region close to the maximum is given by $\tau = - (\ell/v_0)
\ln f$.  Therefore, this adiabatic regime is realized when the 
average time between switches, $k^{-1}$,
is much longer than  $\tau$, which leads to $1> f \gg \exp(-1/\alpha)$. (Note that the
adiabatic regime is not possible unless there is a fixed point and thus $f<1$.)
Only $\alpha \ll 1$ can satisfy this condition.
In this range of the external force, the
mean velocity is simply half the spatial period of the ``substrate"
potential divided by the 
mean switching time, that is, $\langle \dot{x} \rangle = Lk/2 = 2 v_0
\alpha/\Gamma$. 
The  same result can be obtained directly from
Eq.~(\ref{eq:hyper_case2_xdot}) by taking the appropriate
limit.   This result is interesting in that the mean velocity does not
depend on the external force $F$, nor the magnitude $v_0$ of the fluctuating force, 
and is inversely proportional to the
correlation time of the noise.  A higher transition rate reduces the waiting
period at the potential minimum and thus increases the mean velocity. 
However, increasing $k$ 
reduces the range of occurrence of this adiabatic regime.
These results are in perfect agreement with the computer
simulations (Fig.~\ref{vfhyper}).  The transport  
in the adiabatic
regime has been  
investigated as hypersensitive response in Ref.~\cite{ginzburg}
but without a full analytic solution.

(iv) {\it Hyper-Nonlinear Regime:} For a small external force,
$f < \exp(-1/\alpha)$, the particles manage to advance to the next potential
minimum only when the dichotomous noise realizes exponentially rare cases where it
keeps the same value for times much longer than the correlation time.
Therefore, the mean velocity rapidly falls to zero as $f$ decreases.  Indeed,
from Eq.~(\ref{eq:hyper_case2_xdot}), taking the limit $f \rightarrow 0$,
we find for the mean velocity 
\begin{equation}
\langle\dot{x}\rangle/v_0\approx
[2\alpha\Gamma (\ln f)^2]^{-1}.
\end{equation}

This result is striking in that the susceptibility (or the conductance) of the system 
$ d \langle\dot{x}\rangle /d f$ diverges at
$f=0$, indicating that the mean velocity is extremely sensitive to the
external force.  Another interesting feature is that the mean velocity is
inversely 
proportional to the transition rate $k$, in contrast to the adiabatic
regime where the velocity is directly proportional to the transition rate.
Figure~\ref{vfhyper}
shows the mean velocity as a function of the applied force $f$ 
for different values of $\alpha$. Note the linear response regime I for all values of $\alpha$; 
the linear response regime II for $\alpha=1$ (practically for all $f$, except the small-$f$ region);
the adiabatic regime for $\alpha=0.01$ and (more restricted) for $\alpha=0.1$; 
and the hyper-nonlinear regime for all $\alpha$.
The nonmonotonic dependence of the mean velocity on $\alpha$ 
(e.g., $\sim
\alpha$ in the adiabatic regime, $\sim \alpha^{-1}$ in the hyper-nonlinear
regime, and $\alpha$-independent in the linear response regime) leads to the
fact that the curves in Fig.~\ref{vfhyper} cross each other in the 
$f$-regions where there is a switch between these various regimes.

In Fig.~\ref{pxhyper} we present two typical profiles of the
probability density $P(x)$ for the two regimes with and without fixed points,
with an abrupt change in shape between these regimes.

\section{Conclusions}
\label{conclusions}
We have derived explicit results for the probability density and drift
velocity in systems that model Stokes' drift and hypersensitive
response driven by dichotomous noise.  We include the situation in which
the asymptotic dynamics crosses unstable fixed points, a case where the
standard 
approaches are not applicable in a straightforward
way~\cite{paper1,paper2}.  We have presented
analytic results for particular choices of potentials and parameters
that elucidate the behavior in a way not entirely possible when
only numerical results are available.  
 
\appendix

\section{Stokes' Drift Problem: Results for the Piecewise Linear 
Wave Eq.~(\ref{stokes16})}
\label{stokesap}

In the absence of fixed points one finds the following general
piecewise expression for the probability density in four different regions:

\begin{equation}
P(y) =
\begin{cases}
   \displaystyle\frac{\langle\dot{y}\rangle}
                     {L}
   \left[
        C_1\, e^{2 \phi_1 y/L}-\displaystyle\frac{1}{F_-}
   \right]
&  \text{for } y \in [0,L/2-2l) \\[18pt]
   \begin{aligned}
      &
      \displaystyle\frac{\langle\dot{y}\rangle}
                        {L}
      \displaystyle\frac{\text{sgn}[F^2(y)-A^2]}
                        {\left|
                              F^2(y)-A^2
                         \right|^{1-\alpha}} \\[4pt]
      &\times
      \left[
           \frac{(2\alpha-1)k}{\alpha}
           \displaystyle\int_{L/2-2l}^y
           \displaystyle\frac{dz}
                             {\left|
                                   F^2(z)-A^2
                              \right|^{\alpha}}
           +C_2
      \right]
   \end{aligned}
&  \text{for } y \in [L/2-2l,L/2) \\[36pt]
 \displaystyle\frac{\langle\dot{y}\rangle}
                     {L}
   \left[
      C_3\, e^{\phi_2 (2y/L-1)} - \displaystyle\frac{1}{F_+}
   \right] 
&  \text{for } y \in [L/2,L-2l) \\[18pt]
   \begin{aligned}
      &
      \displaystyle\frac{\langle\dot{y}\rangle}
                        {L}
      \displaystyle\frac{\text{sgn}[F^2(y)-A^2]}
                        {\left|
                              F^2(y)-A^2
                         \right|^{1+\alpha}} \\[4pt]
      &\times
     \left[ 
          \frac{(2\alpha+1)k}{\alpha} \displaystyle\int_{L-2l}^y dz\;
          \left|
               F^2(z)-A^2
          \right|^{\alpha}
          +C_4
     \right]
   \end{aligned}
&  \text{for } y \in [L-2l,L)
\end{cases}
\label{generalPy}
\end{equation}
where we have used the dimensionless quantities
$\alpha=lk/bv$, 
and ${\phi_1}$, ${\phi_2}$ as defined in Eq.~(\ref{phi12}).
Note that the system is invariant under the transformation,
$y\rightarrow (y-L/2)$ and $b\rightarrow -b$. Therefore, 
$P(y)$ for $y\,>\,L/2$ can formally be
obtained from the expression for
$P(y-L/2)$ by simply changing $b$ to $-b$ and ({\em nota bene})
$\alpha$ to $-\alpha$.
The integration constants $C_1$--$C_4$ assure that  $P(y)$ has no
discontinuity in this case and is periodic. The mean asymptotic velocity
$\langle\dot{x}\rangle = v + \langle\dot{y}\rangle$ is obtained by
imposing the normalization of $P(y)$.

We turn now to the case of principal interest 
to us, $F_{-}<A<F_{+}$,
where the dynamics exhibits two fixed
points, one stable,
$y_1=L/2-2l+l\left[A-F_{-}\right]/(bv)$, and one unstable,
$y_2=L-2l+l\left[F_{+}-A\right]/(bv)$.
The piecewise expression for the probability density is now given by
\begin{equation}
P(y)=
\begin{cases}
   \displaystyle\frac{\langle\dot{y}\rangle}
                     {L}
   \left[
         D_1 e^{2 \phi_1 y/L}-\displaystyle\frac{1}{F_-}
   \right]
&  \text{for } y\in [0,L/2-2l) \\[18pt]
   \begin{aligned}
      \displaystyle\frac{\langle\dot{y}\rangle}
                        {L}
      &
      \displaystyle\frac{1}
                        {\left[A^2-F^2(y)\right]^{1-\alpha}}
          \left\{
           D_2 - \displaystyle\frac{k(2\alpha-1)}
                                   {\alpha}
      \right. \\[4pt]
      &\times   
      \left.
          \displaystyle \int_{L/2-2l}^{y}dz\;
          \left[F^2(z)-A^2\right]^{-\alpha}
      \right\}
   \end{aligned} 
&  \text{for } y \in [L/2-2l,y_1) \\[36pt]
   \begin{aligned}
      \displaystyle\frac{\langle\dot{y}\rangle}
                        {L}
      &
      \displaystyle\frac{1}
                        {\left[F^2(y)-A^2\right]^{1-\alpha}}
      \left\{
          D_3-\frac{k(2\alpha-1)}
                    {\alpha} 
      \right. \\[4pt]
      & \times  
      \left.   
          \displaystyle\int_{y}^{L/2}dz\;
          \left[A^2-F^2(z)\right]^{-\alpha}
      \right\}
   \end{aligned}
&  \text{for } y \in [y_1,L/2) \\[36pt]
   \displaystyle\frac{\langle\dot{y}\rangle}
                    {L}
   \left\{
        D_4\, e^{2 \phi_2(y/L-1/2)} -\displaystyle\frac{1}{F_+}
   \right\}
&  \text{for } y \in [L/2,L-2l) \\[18pt]
   \begin{aligned}
       \displaystyle\frac{\langle\dot{y}\rangle}
                         {L}
       &
       \displaystyle\frac{k(2\alpha+1)} 
                         {\alpha\;\left| F^2(y)-A^2 \right|^{1+\alpha}}\\[4pt]
       &\times 
       \int_{y_2}^{y}dz\;\text{sgn}\left[F^2(z)-A^2\right]
       |F^2(z)-A^2|^{\alpha}
   \end{aligned}
&  \text{for } y\in [L-2l,L) 
\end{cases}
.
\label{stokes36}
\end{equation}
The integration constants $D_1$--$D_4$ are obtained by imposing the
continuity of $P(y)$ at $y=L/2-2l$, $L/2$, and $L-2l$, and its periodicity.
Note from Eq.~(\ref{stokes36}) that $P(y)$ is continuous at $y_2$, 
$\lim_{y\nearrow y_2}P(y)=\lim_{y\searrow y_2}P(y)=-
\displaystyle({\langle\dot{y}\rangle}/{L})\displaystyle
{[(2\alpha+1)]}/{[2(\alpha+1)A]}$.
Also, $\langle\dot{y}\rangle<0$ [from the positiveness of $P(y_2)$].
At the stable fixed point, $P(y)$ is either divergent, with a weak,
integrable singularity whenever $\alpha\leqslant1$, or finite,
$\lim_{y\nearrow y_1}P(y)=\lim_{y\searrow y_1}P(y)=-
\displaystyle({\langle\dot{y}\rangle}/{L})\displaystyle
{[(2\alpha-1)]}{[2(\alpha-1)A]}$ when $\alpha>1$.
Finally, the mean velocity is obtained by
imposing the normalization condition for $P(y)$. 

The integrals that appear in the expressions (\ref{generalPy}) and (\ref{stokes36}) of $P(y)$
and, correspondingly,
the integration constants $C_1$--$C_4$ and $D_1$--$D_4$ can be
evaluated in a closed analytic form only for certain values of $\alpha$. 
As an example, we present below these results for the particular
case $\alpha=1/2$.

When $A>F_{+}$, the probability density reads
\begin{equation}
P(y)=
\begin{cases}
   \hspace{8pt}\displaystyle\frac{\langle\dot{y}\rangle}
                      {L}
    \left[
         C_1\, e^{2{\phi_1}y/L}-\displaystyle\frac{1}
                                                {F_-}
    \right]
&   \text{for } y \in [0,L/2-2l)\\[12pt]
   -\displaystyle\frac{\langle\dot{y}\rangle}
                      {L}
    \frac{C_2}
         {\left[
               A^2-F^2(y)
          \right]^{1/2}}
&  \text{for } y \in [L/2-2l,L/2)\\[12pt]
  \hspace{8pt}\displaystyle\frac{\langle\dot{y}\rangle}
                     {L}
   \left[
        C_3\, e^{\phi_2 (2y/L-1)}
       -\displaystyle\frac{1}
                          {F_+}
   \right]
&  \text{for } y \in [L/2,L-2l)\\[12pt]
  -\displaystyle\frac{\langle\dot{y}\rangle}
                     {L}
   \left\{ 
        \displaystyle\frac{F(y)}
                          {\left[
                                A^2-F^2(y)
                           \right]}
       +\displaystyle\frac{
                          A^2 \mbox{arcsin} [F(y)/A]+C_4}
                          {\left[
                                A^2-F^2(y)
                           \right]^{3/2}}
   \right\}
&  \text{for } y \in [L-2l,L)
\end{cases}
\label{stokes24}
\end{equation}
and the corresponding asymptotic velocity 
\begin{equation}
   \begin{split}
       \langle\dot{x}\rangle
       =& v + \left\{
                   C_1\displaystyle\frac{e^{\phi_1(1-\Gamma)}-1}
                                        {2\phi_1}
                  +C_3\displaystyle\frac{e^{\phi_2(1-\Gamma)}-1}
                                        {2\phi_2}
                  -\displaystyle\frac{v(1-\Gamma)}
                                     {F_+ F_-}
              \right. \\
       &+     \left.
                   \displaystyle\frac{\Gamma}{4bv}
                   \left[
                        \left(
                              C_2-\displaystyle\frac{F_-}
                                                    {\sqrt{A^2-F_-^2}}
                        \right)
                        \mbox{arcsin}\left(
                                      \displaystyle\frac{F_-}{A}
                                 \right) 
                   \right.
              \right. \\
       &-     \left. 
                   \left.
                       \left(
                             C_2-\displaystyle\frac{F_+}
                                                   {\sqrt{A^2-F_+^2}}
                        \right)
                        \mbox{arcsin}\left(
                                      \displaystyle\frac{F_+}{A}
                                 \right)
		     \right.
		\right. \\
	&+     \left.
                  \left.
                    +\displaystyle\frac{C_4}{A^2}\left(
                        \displaystyle\frac{F_-}{\sqrt{A^2-F_-^2}}
                       -\displaystyle\frac{F_+}
                                          {\sqrt{A^2-F_+^2}}
                                                   \right)
                  \right]
              \right\}^{-1} ,
    \end{split} 
\end{equation}
where $\Gamma = \displaystyle\frac{4l}{L}$ $(0<\Gamma <1)$, and 
$\phi_{1,2}=\displaystyle\frac{2bF_\mp^2}{\Gamma(F_\mp^2-A^2)}$.
The integration constants $C_1$--$C_4$ are given by
\begin{multline}
   \begin{aligned}
       C_1 =&
       \left\{
            \displaystyle\frac{A^2-F_-^2}{A^2-F_+^2}
            -e^{(\phi_1+\phi_2)(1-\Gamma)}
       \right\}^{-1} 
       \left\{
            \displaystyle\frac{A^2-F_-^2}
                              {F_-(A^2-F_+^2)}
           -\displaystyle\frac{\sqrt{A^2-F_+^2}}
                              {F_+\sqrt{A^2-F_-^2}}
       \right.  \\
    &-
       \left.
           \displaystyle\frac{F_+\sqrt{A^2-F_+^2}
                             -F_-\sqrt{A^2-F_-^2}
                             -A^2[\mbox{arcsin}(F_+/A)-\mbox{arcsin}(F_-/A)]}
                             {(A^2-F_+^2)\sqrt{A^2-F_-^2}}
        \right. \\
    &+
       \left.
            e^{\phi_2(1-\Gamma)}
            \left[
                 \displaystyle\frac{\sqrt{A^2-F_+^2}}
                                   {F_+\sqrt{A^2-F_-^2}}
                -\displaystyle\frac{1}{F_-}
            \right]
        \right\}, 
   \end{aligned}
 \\
\shoveleft{
   \begin{aligned}
      C_2 =&
      \left\{ 
           \displaystyle\frac{A^2-F_-^2}{A^2-F_+^2}
           -e^{(\phi_1+\phi_2)(1-\Gamma)}
      \right\}^{-1} 
      \left\{
            \displaystyle\frac{(A^2-F_-^2)^{3/2}}
                              {(A^2-F_+^2)F_-}
            +e^{\phi_1(1-\Gamma)}
       \right.  \\
   &\times
       \left.
            \left\{
                 \displaystyle\frac{\sqrt{A^2-F_+^2}}
                                   {F_+}
                -\displaystyle\frac{(A^2-F_-^2)^{3/2}}
                                   {(A^2-F_+^2)F_-}
                -\displaystyle\frac{\sqrt{A^2-F_+^2}
                                    e^{\phi_2(1-\Gamma)}}
                                   {F_+}
            \right.
       \right.   \\
    &+ 
       \left.
           \left. 
               \displaystyle\frac{F_+\sqrt{A^2-F_+^2}
                                 -F_-\sqrt{A^2-F_-^2}
                                 +A^2[\mbox{arcsin}(F_+/A)
                                     -\mbox{arcsin}(F_-/A)]}
                                  {A^2-F_+^2}
           \right \} 
      \right \},
   \end{aligned}
}
\\
\shoveleft{
   \begin{aligned}
       C_3 =& 
       \left\{
            \displaystyle\frac{A^2-F_-^2}{A^2-F_+^2}
            -e^{(\phi_1+\phi_2)(1-\Gamma)}
       \right\}^{-1}
       \left\{
            \displaystyle\frac{A^2-F_-^2}
                              {F_+(A^2-F_+^2)}
           -\displaystyle\frac{(A^2-F_-^2)^{3/2}}
                              {F_-(A^2-F_+^2)^{3/2}}
       \right.  \\
    &-
       \left.
            \left[
                 \displaystyle\frac{F_+\sqrt{A^2-F_+^2}
                                   -F_-\sqrt{A^2-F_-^2}
                                   +A^2[\mbox{arcsin}(F_+/A)
                                       -\mbox{arcsin}(F_-/A)]}
                                   {(A^2-F_+^2)^{3/2}}
            \right.
       \right.   \\
   &-
       \left.
            \left.
                \displaystyle\frac{(A^2-F_-^2)^{3/2}}
                                  {F_-(A^2-F_+^2)^{3/2}}
               +\displaystyle\frac{1}{F_+}
            \right]
            e^{\phi_1(1-\Gamma)}
       \right\},
   \end{aligned}
}
\\
\shoveleft{
   \begin{aligned}
       C_4 =&
       \left\{
            \displaystyle\frac{A^2-F_-^2}{A^2-F_+^2}
            -e^{(\phi_1+\phi_2)(1-\Gamma)}
       \right\}^{-1}
       \left\{
            \displaystyle\frac{\sqrt{A^2-F_+^2}(A^2-F_-^2)}
                              {F_+}
       \right. \\
   &+
       \left.
           \displaystyle\frac{(A^2-F_-^2)F_+\sqrt{A^2-F_+^2}
                              +A^2\mbox{arcsin}(F_+/A)]}
                             {A^2-F_+^2}
       \right.  \\
   &+
       \left.
            e^{\phi_2(1-\Gamma)}
            \left[
                 \displaystyle\frac{(A^2-F_-^2)^{3/2}}
                                   {F_-}
                -\displaystyle\frac{\sqrt{A^2-F_+^2}(A^2-F_-^2)}
                                   {F_+}
            \right]
       \right.  \\
   &-
       \left.
            e^{(\phi_1+\phi_2)(1-\Gamma)} 
            \left[
                F_-\sqrt{A^2-F_-^2}+A^2\mbox{arcsin}(F_-/A)
               +\displaystyle\frac{(A^2-F_-^2)^{3/2}}
                                  {F_-} 
            \right] 
       \right\} .
   \end{aligned}
}
\end{multline}

For $0<A<F_-$, one obtains the probability density
\begin{equation}
P(y)=
\begin{cases}
   \displaystyle\frac{\langle\dot{y}\rangle}
                     {L}
   \left[ 
        C_1^\prime e^{2\phi_1 y/L}-\displaystyle\frac{1}{F_-}
   \right]
&  \text{for } y \in [0,L/2-2l) \\[12pt]
   \displaystyle\frac{\langle\dot{y}\rangle}
                     {L}
   \displaystyle\frac{C_2^\prime}
                     {\sqrt{F^2(y)-A^2}}
&  \text{for } y \in [L/2-2l,L/2)\\[12pt]
   \displaystyle\frac{\langle\dot{y}\rangle}
                     {L}
   \left\{
        C_3^\prime e^{\phi_2(2y/L-1)}
        -\displaystyle\frac{1}
                           {F_+}
   \right\}
&  \text{for } y \in [L/2,L-2l)\\[12pt]
   \begin{aligned}
       &
       \displaystyle\frac{\langle\dot{y}\rangle}
                         {L}
       \left\{
             \displaystyle\frac{F(y)}
                               {\left[
                                     F^2(y)-A^2
                                \right]}
       \right.\\[4pt]
       &+
       \left.
             \displaystyle\frac{A^2\ln
                               \left[
                                    \displaystyle\frac{\sqrt{F^2(y)-A^2}-F(y)}
                                               {A}
                               \right]
                               +C_4^\prime}
                               {\left[F^2(y)-A^2\right]^{3/2}}
      \right\}
   \end{aligned}
&  \text{for } y \in [L-2l,L)
\end{cases}
\label{stokes26}
\end{equation}
and the corresponding asymptotic drift velocity
\begin{equation}
   \begin{split}
      \langle\dot{x}\rangle
      =& v + \left\{
                  C_1^\prime \displaystyle\frac{e^{\phi_1(1-\Gamma)}-1}
                                        {2\phi_1}
                 +C_3^\prime \displaystyle\frac{e^{\phi_2(1-\Gamma)}-1}
                                        {2\phi_2}
                 -\displaystyle\frac{v(1-\Gamma)}
                                    {F_+F_-}
             \right. \\
       &+    
             \left.
                  \displaystyle\frac{\Gamma}{4bv}
                  \left[
                       \left(
                            C_2^\prime - \displaystyle\frac{F_+}
                                                    {\sqrt{F_+^2-A^2}}
                       \right)
                       \ln \left(
                                \frac{\sqrt{F_+^2-A^2}+F_+}
                                     {A}
                           \right)
                 \right.
            \right. \\
       &-   
            \left.
                 \left.
                      \left(
                           C_2^\prime - \displaystyle\frac{F_-}
                                                   {\sqrt{F_-^2-A^2}}
                      \right)
                      \ln\left(
                              \frac{\sqrt{F_-^2-A^2}+F_-}
                                   {A}
                         \right)
		      \right.
		   \right. \\
 	&+
		\left.
		   \left.
                     \displaystyle\frac{C_4^\prime}{A^2}\left(
                          \displaystyle\frac{F_-}{\sqrt{F_-^2-A^2}}
                           -\displaystyle\frac{F_+}{\sqrt{F_+^2-A^2}}
                                                         \right)
                 \right]
            \right\}^{-1}
\end{split} .
\end{equation}
The integration constants $C_1^\prime$-$C_4^\prime$ are given by
\begin{multline}
   \begin{aligned}
       C_1^\prime =& 
       \left[
            \displaystyle\frac{F_-^2-A^2}{F_+^2-A^2}
            -e^{(\phi_1+\phi_2)(1-\Gamma)}
       \right]^{-1} 
       \left[
            e^{\phi_2(1-\Gamma)}
            \left[
                 \displaystyle\frac{\sqrt{F_+^2-A^2}}
                                   {F_+\sqrt{F_-^2-A^2}}
                -\displaystyle\frac{1}{F_-}
            \right] 
       \right. \\
   &+
       \left.
           \displaystyle\frac{F_-^2-A^2}
                             {F_-(F_+^2-A^2)}
          -\displaystyle\frac{\sqrt{F_+^2-A^2}}
                             {F_+\sqrt{F_-^2-A^2}}
          +\displaystyle\frac{F_+\sqrt{F_+^2-A^2}
                             -F_-\sqrt{F_-^2-A^2}}
                             {(F_+^2-A^2)\sqrt{F_-^2-A^2}}
       \right. \\
   &+
       \left. 
          \displaystyle\frac{A^2}
                            {(F_+^2-A^2)\sqrt{F_-^2-A^2}}
          \ln \left(
                   \displaystyle\frac{\sqrt{F_-^2-A^2}+F_-}
                                     {\sqrt{F_+^2-A^2}+F_+}
              \right)
       \right] ,
   \end{aligned} 
\\
\shoveleft{
   \begin{aligned}
      C_2^\prime =&
      \left[ 
           \displaystyle\frac{F_-^2-A^2}{F_+^2-A^2}
           -e^{(\phi_1+\phi_2)(1-\Gamma)}
      \right]^{-1} 
      \left\{
            -\displaystyle\frac{(F_-^2-A^2)^{3/2}}
                               {(F_+^2-A^2)F_-}
            +e^{\phi_1(1-\Gamma)}
       \right. \\
   &\times
       \left.
            \left[
                -\displaystyle\frac{\sqrt{F_+^2-A^2}}
                                   {F_+}
                +\displaystyle\frac{(F_-^2-A^2)^{3/2}}
                                   {(F_+^2-A^2)F_-}
                +\displaystyle\frac{\sqrt{F_+^2-A^2}
                                    e^{\phi_2(1-\Gamma)}}
                                   {F_+}
            \right.
       \right.  \\
   &+ 
       \left.
           \left. 
               \displaystyle\frac{F_+\sqrt{F_+^2-A^2}
                                 -F_-\sqrt{F_-^2-A^2}}
                                 {F_+^2-A^2}
              +\displaystyle\frac{A^2}
                                 {F_+^2-A^2}
               \ln  \left(
                         \displaystyle\frac{\sqrt{F_-^2-A^2}+F_-}
                                           {\sqrt{F_+^2-A^2}+F_+}
                    \right)
           \right]
      \right\} ,
   \end{aligned}
}
\\
\shoveleft{
   \begin{aligned}
       C_3^\prime =& 
       \left[
            \displaystyle\frac{F_-^2-A^2}{F_+^2-A^2}
           -e^{(\phi_1+\phi_2)(1-\Gamma)}
       \right]^{-1}
       \left\{
            \displaystyle\frac{F_-^2-A^2}
                              {F_+(F_+^2-A^2)}
           -\displaystyle\frac{(F_-^2-A^2)^{3/2}}
                              {F_-(F_+^2-A^2)^{3/2}}
       \right.  \\
    &+
       \left.
            \left[
                 \displaystyle\frac{F_+\sqrt{F_+^2-A^2}
                                   -F_-\sqrt{F_+^2-A^2}}
                                   {(F_+^2-A^2)^{3/2}}
                +\displaystyle\frac{A^2}
                                   {(F_+^2-A^2)^{3/2}}
               \ln  \left(
                         \displaystyle\frac{\sqrt{F_-^2-A^2}+F_-}
                                           {\sqrt{F_+^2-A^2}+F_+}
                    \right)              
            \right.
       \right.  \\
   &+
       \left.
            \left.
                \displaystyle\frac{(F_-^2-A^2)^{3/2}}
                                  {F_-(F_+^2-A^2)^{3/2}}
               -\displaystyle\frac{1}{F_+}
            \right]
            e^{\phi_1(1-\Gamma)}
       \right\},
   \end{aligned}
}
\\
\shoveleft{
   \begin{aligned}
       C_4^\prime =&
       \left[-
            \displaystyle\frac{F_-^2-A^2}{F_+^2-A^2}
            -e^{(\phi_1+\phi_2)(1-\Gamma)}
       \right]^{-1}
       \left\{
            \displaystyle\frac{\sqrt{F_+^2-A^2}(F_-^2-A^2)}
                              {F_+}
       \right. \\
   &+
       \left.
           \displaystyle\frac{F_-^2-A^2}
                             {F_+^2-A^2}
           \left[
                F_+\sqrt{F_+^2-A^2}-A^2 \ln
                \left(
                     \displaystyle\frac{\sqrt{F_+^2-A^2}+F_+}
                                       {A}
                \right)
           \right]
           +e^{\phi_2(1-\Gamma)}
       \right. \\
   &\times
       \left.
            \left[
                -\displaystyle\frac{(F_-^2-A^2)^{3/2}}
                                   {F_-}
                +\displaystyle\frac{\sqrt{F_+^2-A^2}(F_-^2-A^2)}
                                   {F_+}
            \right]
            -e^{(\phi_1+\phi_2)(1-\Gamma)}
       \right.  \\
   &\times
       \left.
            \left[
                F_-\sqrt{F_-^2-A^2 }- A^2 \ln
                \left(
                     \displaystyle\frac{\sqrt{F_-^2-A^2}+F_-}
                                       {A}
                \right)
               -\displaystyle\frac{(F_-^2-A^2)^{3/2}}
                                  {F_-} 
            \right] 
       \right\}.
   \end{aligned}
}
\end{multline}

For $F_-<A<F_+$, the probability density is expressed as
\begin{equation}
P(y)= 
\begin{cases}
   \displaystyle\frac{\langle\dot{y}\rangle}
                     {L}
   \left( D_1\, e^{2 \phi_1 y/L}-\displaystyle\frac{1}{F_-} \right)
&  \text{for } y\in [0,L/2-2l) \\[12pt]
   \displaystyle\frac{\langle\dot{y}\rangle}
                     {L}
   \displaystyle\frac{D_2}
                     {\sqrt{A^2-F^2(y)}}
&  \text{for } y \in [L/2-2l,y_1) \\[12pt]
   \displaystyle\frac{\langle\dot{y}\rangle}
                    {L}
   \displaystyle\frac{D_3}
                     {\sqrt{F^2(y)-A^2}}
&  \text{for } y \in (y_1,L/2) \\[12pt]
   \displaystyle\frac{\langle\dot{y}\rangle}
                     {L}
   \left[
        D_4\, e^{2\phi_2(y/L-1/2)}-\displaystyle\frac{1}{F_+}
   \right]
&  \text{for } y\in [L/2,L-2l) \\[12pt]
   \begin{aligned}
       \displaystyle\frac{\langle\dot{y}\rangle}
                         {L}
       &
       \left\{
            \displaystyle\frac{F(y)}
                              {\left[F^2(y)-A^2\right]}
       \right. \\
       &+
       \left.
            \displaystyle\frac{A^2\ln
                               \left\{
                                    \left[
                                         \sqrt{F^2(y)-A^2}-F(y)
                                    \right]/A
                               \right\}}
                              {\left[
                                    F^2(y)-A^2
                               \right]^{3/2}} 
       \right\}
   \end{aligned}
&  \text{for } y \in [L-2l,y_2) \\[24pt]
   \begin{aligned}
      \displaystyle\frac{\langle\dot{y}\rangle}
                     {L}
   &
      \left\{
        \displaystyle\frac{-F(y)}
                          {\left[A^2-F^2(y)\right]}
      \right.\\
   &+
      \left.
          \displaystyle\frac{-A^2\left[
                                      \pi/2 + A^2\mbox{arcsin}[F(y)/A]
                                 \right]}
                            {\left[A^2-F^2(y)\right]^{3/2}}
   \right\}   
   \end{aligned}
&  \text{for } y \in (y_2,L)
\end{cases}
\label{stokes40}
\end{equation}  
and the mean drift velocity as
\begin{equation}
    \begin{split}
        \langle\dot{x}\rangle
    =&
        v + \left\{
                 D_1 \displaystyle\frac{e^{\phi_1(1-\Gamma)}-1}
                                       {2\phi_1}
                +D_4 \displaystyle\frac{e^{\phi_2(1-\Gamma)}-1}
                                       {2\phi_2}
                    -\displaystyle\frac{v(1-\Gamma)}
                                       {F_+F_-}
            \right. \\
     &+     
            \left.
            \displaystyle\frac{\Gamma}
                              {4bv}
            \left[
                 \left(
                      D_2+\frac{F_-}{\sqrt{A^2-F_-^2}}
                 \right)
                 \left(
                      \pi/2-\mbox{arcsin} \left(F_-/A\right)
                 \right)
            \right. 
        \right.  \\
     &+     
        \left. 
            \left.
                 \left(
                      D_3 - \displaystyle\frac{F_+}
                                              {\sqrt{F+^2-A^2}}
                 \right)
                 \ln \left(
                           \displaystyle\frac{\sqrt{F_+^2-A^2}+F_+}
                                             {A}
                     \right)
            \right]
        \right\}^{-1}
    \end{split}
\end{equation}
where the integration constants $D_1$--$D_4$ are given by
\begin{multline}
       D_1=\displaystyle\frac{1}
                         {F_-}
       +\displaystyle\frac{F_-\sqrt{A^2-F_-^2}
                          +A^2\mbox{arcsin}(F_-/A)-A^2\pi/2}
                          {\left(A^2-F_{-}^2\right)^{3/2}},
\\   
\shoveleft{
\begin{aligned}
       D_2 
       =&   \displaystyle\frac{\sqrt{A^2-F_-^2}
                               \bigl[
                                    e^{\phi_1(1-\Gamma)}-1
                               \bigr]}
                              {F_-} \\
       &+
            e^{\phi_1(1-\Gamma)}
            \left[
                 \displaystyle\frac{F_-\sqrt{A^2-F_-^2}
                                    +A^2\mbox{arcsin}(F_-/A)-A^2\pi/2}
                                   {A^2-F_-^2}
           \right] , 
\end{aligned} }
\\
\shoveleft{
\begin{aligned}
       D_3
       =&   \displaystyle\frac{\sqrt{F_+^2-A^2}
                               \bigl[
                                     e^{-\phi_2(1-\Gamma)}-1
                               \bigr]}
                              {F_+} \\
        &+  e^{-\phi_2(1-\Gamma)}
            \left\{
                 \displaystyle\frac{A^2\ln
                                    \left[
                                         \left(
                                              \sqrt{F_+^2-A^2}+F_+
                                         \right)/A
                                    \right]
                                    -F_+\sqrt{F_+^2-A^2}}
                                   {F_+^2-A^2}
           \right\}  
\end{aligned}},
\\
\shoveleft{
\begin{aligned}
       D_4= e^{-\phi_2(1-\Gamma)}
       &
       \left\{ 
            \displaystyle\frac{1}
                               {F_+}
       \right. \\
       &+
       \left.
       \displaystyle\frac{A^2\ln
                          \left[
                               \left(
                                    \sqrt{F_+^2-A^2}+F_+
                               \right)/A
                          \right]
                           -F_+\sqrt{F_+^2-A^2}}
                         {\left(F_+^2-A^2\right)^{3/2}}
       \right\}.
\end{aligned}}
\end{multline}

\section{The Problem of Hypersensitive Transport: Particular Cases for the
Piecewise Linear Velocity Profile}
\label{hyperparticular}

Again, explicit results can be obtained for the case $\alpha=1/2$.
When there are no fixed points one obtains for the probability density:
\begin{equation}
P(x)=
\begin{cases}
   \begin{aligned}
       \displaystyle\frac{\langle\dot{x}\rangle}
                         {LF}
&
       \left\{
            1+\displaystyle\frac{2f\Delta\mbox{arcsin}(1/f)}
                                {(\Delta+1)(f+1)^{3/2}\sqrt{f-1}}
       \right.
\\[4pt]
&\times
       \left.
            \exp\left[
                      -\displaystyle\frac{4f}
                                         {(f^2-1)\Gamma}
                     \frac{x}{L}
                \right]
       \right\}
   \end{aligned}
&  \text{for } x\in [0,L/2-2l) \\[32pt]
   \begin{aligned}
       \displaystyle\frac{\langle\dot{x}\rangle}
                         {LF}
&
       \Biggl\{
             1+\displaystyle\frac{\chi}
                                 {(f - \chi)^{3/2}\sqrt{f + \chi}}
       \Biggr.
\\[4pt] 
&\times
       \Biggl. 
             \biggl\{
                   \sqrt{f^2-\chi^2}-\sqrt{f^2-1}
             \biggr. 
       \Biggr.
\\
&+     \Biggl.
             \biggl.
                  f\mbox{arcsin}(\chi/f)
                - f\displaystyle\frac{\Delta-1}
                                     {\Delta+1}
                \mbox{arcsin}(1/f)
            \biggr\}
       \Biggr\}
   \end{aligned}
&  \text{for } x\,\in [L/2-2l,L/2)
\end{cases}
\label{hyper19}
\end{equation}
where $\chi = 2\Gamma-1-x/l$.
The mean velocity is
\begin{equation}
\displaystyle\frac{\langle\dot{x}\rangle}
                 {F}=
\left\{
     1-2\Gamma+2\Gamma\sqrt{f^2-1}\mbox{~arcsin}
     \left(\displaystyle\frac{1}
                             {f}
     \right)
     +\displaystyle\frac{\Gamma f(\Delta-1)}
                        {\Delta+1}
     \left[
          \mbox{arcsin}\left(
                        \displaystyle\frac{1}
                                          {f}
                   \right)
     \right]^2
\right\}^{-1},
\label{hyper21}
\end{equation}
where
\begin{equation}
\Delta=\displaystyle\left(
                         \frac{f+1}
                              {f-1}
                    \right)
\exp\left[
         \displaystyle\frac{2f(1-\Gamma)}
                           {(f^2-1)\Gamma}
    \right].
\end{equation}

When there are two fixed points, $P(x)$ has an 
integrable power-law  singularity
at the stable fixed point:
\begin{equation}
P(x)=
\begin{cases}
    \begin{aligned}
        \displaystyle\frac{\langle\dot{x}\rangle}
                          {LF}
&
        \left\{
             1+\exp\left[
                        \displaystyle\frac{4f}
                                          {\Gamma(1-f^2)}
                        \left(
                             \displaystyle\frac{x}
                                               {L}
                            -\displaystyle\frac{1-\Gamma}
                                               {2}
                        \right)
                   \right] 
        \right. 
\\[2pt] 
&\times
        \left.
             \displaystyle\frac{f\ln\left[
                                        (1+\sqrt{1-f^2})/{f}
                                    \right]
                                -\sqrt{1-f^2}}
                               {(1-f)^{3/2}\sqrt{1+f}}
        \right\}
   \end{aligned}
&   \text{for } x \in [0,L/2-2l) \\[28pt]
   \begin{aligned}
       \displaystyle\frac{\langle\dot{x}\rangle}
                         {LF}
&
       \left\{
            1+ \displaystyle\frac{\chi}
                                 {(\chi-f)^{3/2}
                                   \sqrt{\chi+f}}
       \right. 
\\[2pt]
&\times
       \left. 
            \left\{
                 f\ln\left[
                          \displaystyle\frac{\chi+\sqrt{\chi^2-f^2}}
                                            {f}
                     \right] 
                 -\sqrt{\chi^2-f^2}
            \right\}
       \right\}
    \end{aligned}
&   \text{for } x \in [L/2-2l,x_1) \\[28pt]
   \begin{aligned}
       \displaystyle\frac{\langle\dot{x}\rangle}
                         {LF}
&
       \left\{
            1+\displaystyle\frac{\chi}   
                                {(f-\chi)^{3/2}\sqrt{\chi+f}}
       \right.
\\[2pt]
&\times
       \left.
           \left\{
                \sqrt{f^2-\chi^2}+f
                \left[
                     \mbox{arcsin}\left(
                                   \displaystyle\frac{\chi}
                                                     {f}
                              \right)
                     -\displaystyle\frac{\pi}
                                        {2}
                \right]
           \right\}
       \right\}
   \end{aligned}
&  \text{for } x \in (x_1,x_2) \\[28pt]
   \begin{aligned}
       \displaystyle\frac{\langle\dot{x}\rangle}
                         {LF}
&
       \left\{
            1+\displaystyle\frac{\chi}
                                {(f-\chi)^{3/2}(-\chi-f)^{1/2}}
           \Biggl\{
                \sqrt{\chi^2-f^2}
           \Biggr.
       \right.
\\[2pt]
&-
       \left. 
           \left.
               \sqrt{1-f^2}
               +f\ln\left[
                          \displaystyle\frac{\sqrt{\chi^2-f^2}-\chi)}
                                            {\sqrt{1-f^2}+1} 
                    \right]
           \right.
       \right.
\\[2pt]
&-
       \left.
           \left.
               \exp\left(
                        -\displaystyle\frac{2f(1-\Gamma)}
                                            {\Gamma(1-f^2)}
                    \right)  
                \left(
                     \displaystyle\frac{1+f}
                                       {1-f}
                \right)
           \right.
       \right.
\\[2pt]
&\times
       \left.
           \left. 
                \left[f\ln\left(
                               \displaystyle\frac{1+\sqrt{1-f^2}}
                                                 {f}
                          \right)
                    -\sqrt{1-f^2}
                \right]
           \right\}
       \right\}
   \end{aligned}
&  \text{for } x \in (x_2,L/2).
\end{cases}
\label{hyper...}
\end{equation}

For the mean velocity we find
\begin{equation}
   \begin{split}
       \displaystyle\frac{\langle \dot{x}\rangle}
                         {F} 
    =&
       \left\{
            1+\Gamma\left[
                         f\frac{\pi^2}{4}
                         +\frac{f}{2}
                         -\frac{3}{2}
                         +\frac{1+f}{2}
                         \exp\left(
                                 -\frac{2f(1-\Gamma)}
                                       {\Gamma(1-f^2)}
                             \right)
                    \right]
       \right.
\\[2pt]
&+
       \left.
            \left[
                 1-\exp\left(
                           -\frac{2f(1-\Gamma)}
                                 {\Gamma(1-f^2)}
                       \right)
            \right]
            \frac{\Gamma\sqrt{1+f}}
                 {2f\sqrt{1-f}} 
       \right. 
\\[2pt]
&\times
      \left.
           \left[
                 f\ln\left(
                          \frac{1+\sqrt{1-f^2}}{f}
                     \right)
                 -\sqrt{1-f^2}
            \right]  
     \right.
\\[2pt]
&+
     \left. 
          \frac{\Gamma f}{2} 
          \ln^2\left(
                    \frac{1+\sqrt{1-f^2}}
                         {f}
               \right)
          \left[
                1+\frac{1+f}
                       {1-f}
                \exp\left(
                        -\frac{2f(1-\Gamma)}
                              {\Gamma(1-f^2)}
                    \right)
          \right] 
       \right. 
\\[2pt]
&+
       \left.
            \frac{\Gamma}{2} 
            \ln\left(
                    \frac{1+\sqrt{1-f^2}}
                         {f}
               \right) 
            \left[
                 \sqrt{\frac{1-f}
                           {1+f}}
                -\frac{2f^2}
                      {\sqrt{1-f^2}} 
            \right. 
       \right.
\\[2pt]
&-
       \left. 
           \left. 
               (1+2f)\sqrt{\frac{1+f}
                                   {1-f}}
               \exp\left(
                       -\frac{2f(1-\Gamma)}
                             {\Gamma(1-f^2)}
                   \right)
           \right]
       \right\}^{-1}.
   \end{split}
\label{hyper34}
\end{equation}

\section*{Acknowledgments}
This work was partially supported by the Swiss National Science
Foundation (I.B.) and by the National Science Foundation under grant
Nos. PHY-0354937.


\begin{thebibliography}{100}
\bibitem{footnote} The half delta functions are defined as
\begin{align*}
&\lim_{\varepsilon \searrow 0}\;\int_{-\infty}^{-\varepsilon}
\delta_-(x) f(x) dx =  
(1/2)\lim_{x\nearrow 0}f(x)=(1/2)f(-0)\\
&\lim_{\varepsilon \searrow 0}\;\int_{+\varepsilon}^{+\infty} \delta_-(x) f(x)
dx = 0\\
&\lim_{\varepsilon \searrow 0}\;\int_{-\infty}^{-\varepsilon} \delta_+(x)
f(x) dx = 0\\
&\lim_{\varepsilon \searrow 0}\;\int_{+\varepsilon}^{+\infty} \delta_+(x) f(x)
dx =(1/2)\lim_{x\searrow 0}f(x)=(1/2)f(+0)
\end{align*}
with $\delta(x)=\delta_-(x)+\delta_+(x)$.
\end{thebibliography}
\end{document}